\documentclass[12pt]{article}
\newcommand{\be}{\begin{equation}}
\newcommand{\ee}{\end{equation}}
\newcommand{\bea}{\begin{eqnarray}}
\newcommand{\eea}{\end{eqnarray}}
\newcommand{\sptwo}{1.4}
\newcommand{\doublespace}{\edef\baselinestretch{\sptwo}\Large\normalsize}
\newcommand{\newsection}[1]{
\section{#1}
\setcounter{equation}{0}}

\newcounter{newapp}
\begin{document}
\begin{center}
{\large\bf Non-BPS Brane Dynamics And Duality}
\end{center}
~\\
\begin{center}
{\bf T.E. Clark}\footnote{e-mail address: clark@physics.purdue.edu} and 
{\bf Muneto Nitta}\footnote{e-mail address: nitta@th.phys.titech.ac.jp\\
\hspace*{0.25in}Current address: {\it Department of Physics\\
\hspace*{1.3in}Tokyo Institute of Technology\\
\hspace*{1.3in}Tokyo 152-8551, Japan}}\\
{\it Department of Physics\\
Purdue University\\
West Lafayette, IN 47907-1396}\\
~\\
And\\
~\\
{\bf T. ter Veldhuis\footnote{e-mail address: terveldhuis@macalester.edu}}\\

{\it Department of Physics \& Astronomy\\
Macalester College\\
Saint Paul, MN 55105-1899}
~\\
~\\
\end{center}
\begin{center}
{\bf Abstract}
\end{center}
The Green-Schwarz action for a non-BPS p=2 brane embedded in a N=1, D=4 target 
superspace is shown to be equivalent to the Nambu-Goto-Akulov-Volkov action obtained 
via the nonlinear realization of the associated target space super-Poincar\'{e} 
symmetries.  Introducing a U(1) gauge field strength as a Lagrange multiplier, this 
p=2 brane action is re-cast into its equivalent dual form non-BPS D2-brane 
Born-Infeld action.  Following the procedure given by Sen, the Green-Schwarz action 
for a non-BPS D2-brane is determined directly.  From it, conversely, the dual form non-BPS 
p=2 brane action is derived.  The p=2 brane and the D2-brane actions obtained by these two
approaches are different in form.  Through explicitly determined field redefinitions, these
actions are shown to be equivalent.  
~\\

\newpage
\doublespace

\newsection{Introduction}

Super p-brane dynamics can be obtained by means of the 
Green-Schwarz method \cite{GS} for constructing the brane action.  In the BPS 
case the action consists of two sets of separately super-Poincar\'e invariant terms 
which are related to each other by $\kappa$-symmetry \cite{Siegel}.  Exploiting 
this local fermionic symmetry, half of the fermionic world volume fields can be set to 
zero along with the Wess-Zumino set of action terms in order to achieve the final form 
of the BPS brane action.  The super Dp-brane actions can be obtained by means of 
duality transformations of the super p-brane actions \cite{Townsend:1995af}.  Indeed the action 
for a BPS super D2-brane embedded in D=10 superspace was first obtained by a 
duality transformation on the action for a BPS super p=2 brane action embedded 
in D=11 superspace \cite{Duff:1992hu}. In addition, non-BPS branes occur in field 
theory \cite{Chibisov:1997rc} as well as string theory \cite{Sen:1999md}\cite{More}\footnote{In 
string theory, unstable non-BPS branes whose effective action includes a tachyon \cite{Tachyon}, 
as well as stable non-BPS branes, are discussed}.  
Excluding the Wess-Zumino action terms from the total Green-Schwarz action ab initio eliminates the 
$\kappa$-symmetry of the action and so prevents the gauging away of half of the fermionic 
fields.  With all fermionic fields present the resulting super-Poincar\'e 
invariant action describes a non-BPS brane action.  This approach can be 
applied to the construction of non-BPS p-brane actions as well as non-BPS Dp-brane actions.   

Besides the above Green-Schwarz approaches to brane dynamics, the coset method \cite{Volkov:jx}\cite{Coleman:sm}
has been used to construct actions for static gauge BPS super p-branes \cite{Ivanov:1999fw} and 
BPS super Dp-branes \cite{Bagger:1996wp}.  Nonlinearly realized supersymmetry in non-BPS D-branes was emphasized in 
\cite{Hara:2000sg}.  In reference \cite{Clark:2002bh} the action for a 
non-BPS p=2 brane embedded in N=1, D=4 superspace was constructed via the nonlinear realization of the spontaneously broken super-Poincar\'e symmetries of the target superspace.  The action described the motion of the brane 
in N=1, D=4 superspace through the brane localized Nambu-Goldstone boson field $\phi$ associated 
with motions in space directions transverse to the brane, hence in the direction 
of the broken space translation symmetry.  It also involved brane localized D=3 Majorana Goldstino 
fields $\theta_i$ and $\lambda_i$, $i=1,2$, describing brane oscillations in Grassmann directions of 
superspace which are associated with the completely broken N=1, D=4 
supersymmetry (SUSY).  The action, after application of the \lq\lq inverse Higgs 
mechanism" \cite{Ivanov:1975zq}, is the N=1, D=4 super-Poincar\'e invariant synthesis of 
the Akulov-Volkov \cite{Volkov:jx} and Nambu-Goto \cite{Nambu:1974zg} actions
\be
\Gamma = -\sigma \int d^3 x \det{\hat{e}} \sqrt{ 1- \lbrack \hat{e}_a^{-
1m}(\partial_m \phi + \bar\theta {\stackrel{\leftrightarrow}{\partial}}_m 
\lambda )\rbrack^2} ,
\label{CNtVAVNG}
\ee
with $\hat{e}_m^{~a}$ the Akulov-Volkov dreibein in the static gauge 
$\hat{e}_m^{~a} = \delta_m^{~a} + i\bar\theta \gamma^a \partial_m \theta 
+i\bar\lambda \gamma^a \partial_m \lambda .$  Exploiting the coset method 
further, the action was shown to be dual to the action for a space-filling 
non-BPS D2-brane whose supersymmetric Born-Infeld \cite{Born:gh} action was given by
\be
\Gamma = -\sigma \int d^3 x \left\lbrack \sqrt{\det{(\hat{g}_{mn} +F_{mn} )}} -
F^m (\bar\theta {\stackrel{\leftrightarrow}{\partial}}_m \lambda )\right\rbrack 
,
\label{CNtVBI}
\ee
where the Akulov-Volkov metric is given by $\hat{g}_{mn} = 
\hat{e}_m^{~a}\eta_{ab}\hat{e}_n^{~b}$ and the field strength vector and tensor are related as $F^m = \frac{1}{2}\epsilon^{mnr}F_{nr}$ where $F_{mn} = \partial_m A_n -\partial_n A_m$ with $A_m$ the Born-Infeld U(1) gauge field.

The purpose of this paper is to determine the action for a non-BPS p=2 brane 
embedded in N=1, D=4 superspace and the action for a space-filling N=2, D=3 non-BPS D2-brane directly by means of the Green-Schwarz \cite{GS} approach as modified by Sen \cite{Sen:1999md} for the non-BPS case.  The actions dual to these actions are then determined.  In the cases that the actions obtained are of a different form than the corresponding brane actions obtained by means of the coset method their equivalence is established through explicit field redefinitions.  
Specifically in section 2, the non-BPS Green-Schwarz method is used to obtain the action for a non-BPS p=2 brane.  Its form, in the static gauge, is the same as that obtained via the coset method, equation (\ref{CNtVAVNG}).  Introducing
a U(1) gauge field strength as a Lagrange multiplier, the dual form of this action is found to have the same Born-Infeld form as obtained using the coset method, equation (\ref{CNtVBI}).  In section 3, the procedure of Sen is applied to the space-filling N=2, D=3 non-BPS D2-brane case in order to obtain its Green-Schwarz action directly.  
This non-BPS D2-brane action in an arbitrary gauge is found to be 
\be
\Gamma = -\sigma \int d^3 \xi \sqrt{\det{(\hat{g}_{mn}} +{\cal{F}}_{mn})} ,
\label{GSSD2}
\ee
where ${\cal F}_{mn} = F_{mn} -b_{mn}$ with the two form $b_{mn}$ given in terms of the brane degrees of freedom so that ${\cal F}_{mn}$ is supersymmetric.  Further this action is shown to be dual to the action for a non-BPS p=2 brane embedded in N=1, D=4 superspace with the form
\be
\Gamma = -\sigma \int d^3 \xi \left\lbrack \det{\hat{e}} \sqrt{ 1 -(\hat{e}_a^{-
1m} \partial_m \phi )^2} -\frac{1}{2} \epsilon^{mnr} b_{mn} \partial_r \phi 
\right\rbrack ,
\label{GSSdualp2}
\ee
where now the two form $b_{mn}$ couples directly to the trivially conserved current 
$\epsilon^{mnr} \partial_r \phi$.  The non-BPS D2-brane action and its dual p=2 brane action obtained by the
modified Green-Schwarz method of Sen, equations (\ref{GSSD2}) and (\ref{GSSdualp2}), respectively, have different forms than the corresponding non-BPS brane actions derived via the coset method (and Green-Schwarz p-brane method), equations (\ref{CNtVBI}) and (\ref{CNtVAVNG}), respectively.  In section 4, the explicit field redefinitions are given that render the two forms of the actions equivalent. 
\pagebreak

\newsection{The Non-BPS p=2 Brane Action And Duality}

The N=1, D=4 superspace coordinates of an embedded p=2 brane are denoted by the 
bosonic space-time coordinates, $x^\mu (\xi^m)$, with $\mu = 0, 1, 2, 3$, and the fermionic Weyl spinor 
coordinates, $\theta_\alpha (\xi^m)$ and $\bar\theta_{\dot\alpha} (\xi^m)$, 
where $\xi^m$, $m=0, 1, 2$, express the world volume space-time coordinates of 
the embedded brane.  The linear representation of the N=1, D=4 super-Poincar\'e 
symmetries on the superspace coordinates is most easily obtained from the motion that group 
multiplication induces on the parameter space of the SUSY graded Lie group.  The N=1, 
D=4 super-translation subgroup elements can be written as
\be
\Omega (x, \theta, \bar\theta) = e^{i(x^\mu P_\mu + \theta^\alpha Q_\alpha + 
\bar\theta_{\dot\alpha} \bar{Q}^{\dot\alpha})}.
\ee
Multiplication by the group elements and use of the Baker-Campbell-Hausdorf 
formula yields the induced motion in superspace.  For translations and SUSY 
transformations it is obtained that the group multiplication law
\be
\Omega (a, \eta, \bar\eta) \Omega (x, \theta, \bar\theta)= \Omega (x+a+i(\eta 
\sigma \bar\theta -\theta \sigma \bar\eta), \theta +\eta, \bar\theta + 
\bar\eta)
\ee
induces a movement in superspace $(x, \theta, \bar\theta)\longrightarrow (x^\prime, \theta^\prime, \bar\theta^\prime)=
(x+a+ i(\eta \sigma \bar\theta -\theta \sigma \bar\eta), \theta +\eta, \bar\theta + \bar\eta)$.  Proceeding similarly, all the N=1, D=4 super-Poincar\'{e} symmetry transformations, including $R$ symmetry (cf. equation (\ref{N2D3})),
are obtained
\begin{center}
\begin{tabular}{ll}
$\delta^P(a) x^\mu = a^\mu$ & $\delta^Q (\eta, \bar\eta) x^\mu = i(\eta 
\sigma^\mu 
\bar\theta - \theta \sigma^\mu \bar\eta)$ \\
$\delta^P(a) \theta_\alpha = 0$ & $\delta^Q (\eta, \bar\eta) \theta_\alpha = 
\eta_\alpha$ \\
$\delta^P(a) \bar\theta_{\dot\alpha} = 0$ & $\delta^Q (\eta, \bar\eta) 
\bar\theta_{\dot\alpha} = \bar\eta_{\dot\alpha}$ \\
 & \\
$\delta^M(\omega) x^\mu = \omega^{\mu\nu} x_\nu$ & $\delta^R(\rho) x^\mu = 0$ \\
$\delta^M(\omega) \theta_\alpha = -\frac{i}{4} \omega^{\mu\nu} (\theta 
\sigma_{\mu\nu})_\alpha$ & $\delta^R(\rho) \theta_\alpha = +i\rho \theta_\alpha$ 
\\
$\delta^M(\omega) \bar\theta_{\dot\alpha} = -\frac{i}{4} \omega^{\mu\nu} 
(\bar\theta \bar\sigma_{\mu\nu})_{\dot\alpha}$ & $\delta^R(\rho) 
\bar\theta_{\dot\alpha} = -i\rho \bar\theta_{\dot\alpha},$ \\
\end{tabular}
\end{center}
\be
~
\ee
where $\delta x^\mu = x^{\prime \mu} - x^\mu$ , and so on.  These variations represent the SUSY charge algebra, for example
\be
[\delta^Q(\eta, \bar\eta) , \delta^Q (\zeta, \bar\zeta)] = +2 \delta^P ( i(\zeta 
\sigma \bar\eta -\eta \sigma \bar\zeta)).
\ee

According to the Green-Schwarz construction \cite{GS}, 
the SUSY (and translation) invariant one-forms $\partial_m \theta$ , $\partial_m 
\bar\theta$ and
\be
\Pi^\mu_m = \partial_m x^\mu +i \partial_m\theta \sigma^\mu \bar\theta - i\theta 
\sigma^\mu \partial_m \bar\theta 
\label{1formsN1D4}
\ee
can be used to construct the translation and SUSY invariant induced world volume 
metric $g_{mn}$
\be
g_{mn} = \Pi^\mu_m \eta_{\mu\nu} \Pi^\nu_n .
\label{metricD4}
\ee
Thus, the action for a non-BPS p=2 brane embedded in N=1, D=4 superspace is 
given by
\be
\Gamma = -\sigma \int d^3 \xi \sqrt{\det{g_{mn}}} ,
\label{Action1}
\ee
where $\sigma$ is the brane tension.  Besides being N=1, D=4 super-Poincar\'{e} 
invariant, the action is world volume reparametrization invariant.  Consider the
reparametrization $\xi = \xi (\xi^\prime)$, the world volume element is rescaled 
by the Jacobian $d^3 \xi = (\det{\partial \xi /\partial \xi^\prime}) d^3 
\xi^\prime$ while the world volume one-form transforms as a world volume vector
$\Pi^\mu_m (\xi) = (\partial \xi^{\prime n} /\partial \xi^m )\Pi^\mu_n 
(\xi^\prime)$.  Hence the metric is indeed a world volume two tensor and 
$\sqrt{\det{g}}$ a world volume density, $\sqrt{\det{g(\xi)}} = 
(\det{\partial \xi^\prime /\partial \xi}) \sqrt{\det{g(\xi^\prime)}}$.  Thus the 
action is left invariant.

The reparametrization invariance can be used to choose the static gauge in which 
the world volume coordinates are chosen to be the first three space-time 
coordinates, $\xi^m = x^m $ while the remaining space point is re-labeled as the 
p=2 brane degree of freedom $\phi (x^m) = x^3 (x^m)$, a world volume field.  
Having chosen the space-time coordinates thusly, it is important to choose a 
commensurate linear combination of fermion fields to describe the brane 
oscillations into Grassmann directions.  More specifically, this choice of gauge 
masks the initial explicit four dimensional symmetry, indeed it leaves explicit 
a three dimensional symmetry.  Exploiting this manifest covariance, the N=1, D=4 
super-Poincar\'{e} algebra can be re-expressed as a centrally extended N=2, D=3 
super-Poincar\'{e} algebra including the D=4 Lorentz symmetry generators as 
additional N=2, D=3 automorphism charges.  The components of the Weyl fermion 
supersymmetry charges $Q_\alpha$ and $\bar{Q}_{\dot\alpha}$ will then be 
combined to form the components of the two N=2, D=3 Majorana spinor 
supersymmetry charges, denoted $q_i$ and $s_i$, $i=1,2$ such that they anti-
commute with themselves to yield the translation generator $p^m =P^m$ and with 
each other to yield the central charge $Z = P_3$, the direction of brane oscillations 
in this gauge.  This rearrangement of components is also reflected in the fermion 
fields.  The Weyl fields will now be written in terms of two D=3 Majorana 
fermion world volume fields, $\theta_i (x^m)$ and $\lambda_i (x^m)$.  
The corresponding relation between the charges and fields are obtained from the equality of the SUSY 
transformations in each basis, that is $\theta^\alpha Q_\alpha + \bar\theta_{\dot\alpha} \bar{Q}^{\dot\alpha} = \bar\theta_i q_i + \bar\lambda_i s_i .$

The action (\ref{Action1}) can be written in terms of these new field variables, $(\phi, \theta, \lambda)$, or equivalently, one can start with the N=2, D=3 SUSY transformations 
and construct the action according to the Green-Schwarz method 
directly.  Implementation of the static gauge will then be straightforward.  Both approaches yield the same expressions for $\Pi_m^\mu$ and hence the action.  Their interrelation will be noted below.  To find the new fields and charges consider the space-time translation generator $P^\mu$, which transforms as a vector
$(\frac{1}{2}, \frac{1}{2})$ representation of the D=4 Lorentz group, it
consists of a D=3 Lorentz group vector, $p^m = P^m$, with $m=0,1,2$, and a D=3 
scalar, $Z\equiv P_3$.  Likewise, the Lorentz transformation charges 
$M^{\mu\nu}$ are in the D=4 $(1,1)_A$ representation which consists of two D=3 
vector representations: $M^{mn}=\epsilon^{mnr} M_r$ and $K^m \equiv M^{m3}$.
The $R$ charge is a singlet from both points of view.  Finally the D=4
SUSY $(\frac{1}{2}, 0)$ spinor $Q_\alpha$ and the $(0,\frac{1}{2})$ spinor
$\bar{Q}_{\dot\alpha}$ consist of two D=3 two-component Majorana spinors: 
$q_i$ and $s_i$, with $i=1,2$, comprising the charges for N=2, D=3 SUSY.  These 
spinors are given as linear combinations of $Q_\alpha$ and 
$\bar{Q}_{\dot\alpha}$ according to
\be
\pmatrix{
q_i \cr
s_i \cr} = \frac{1}{2} e^{i\frac{\pi}{4}} \pmatrix{
\sigma & i\sigma\sigma_z \cr
-i\sigma & -\sigma\sigma_z \cr} \pmatrix{Q_\alpha \cr
\bar{Q}^{\dot\alpha} \cr} ,
\ee
where $(\sigma_x, \sigma_y, \sigma_z)$ are the Pauli matrices and 
\be
\sigma \equiv \pmatrix{
1 & -1 \cr
-i & -i \cr}.
\ee

The N=1, D=4 super-Poincar\'{e} algebra can be written in terms of 
these N=2, D=3 representation charges as
\begin{center}
\begin{tabular}{ll}
$[p^m , p^n ] = 0$ & $[M^m ,M^n ] = -i\epsilon^{mnr} M_r$ \\
$[p^m , Z ] = 0$ & $[M^m ,K^n ] = -i\epsilon^{mnr} K_r $ \\
& $[K^m , K^n ] = +i\epsilon^{mnr} M_r$ \\
& \\
$[M^{m}, p^n ] = -i\epsilon^{mnr} p_r$ & $[K^m , p^n ] = +i\eta^{mn} Z$ \\
$[M^{m} , Z ] = 0$ & $[K^m , Z ] = + i p^m$\\
& \\
$[M^{mn} , q_i ] = -\frac{1}{2} \gamma^{mn}_{ij} q_j\;\;\;\;\;\;\;\;\;$ &
$[K^m , q_i ] = +\frac{1}{2} \gamma^m_{ij} s_j$\\
$[M^{mn} , s_i ] = -\frac{1}{2} \gamma^{mn}_{ij} s_j$  &
$[K^m , s_i ] = -\frac{1}{2} \gamma^m_{ij} q_j$\\
& \\
$[R , q_i ] = +i s_i$ &
$\{ q_i , q_j \} = +2\left(\gamma^m C \right)_{ij} p_m$ \\
$[R , s_i ] = -i q_i$ &
$\{ s_i , s_j \} = +2\left(\gamma^m C \right)_{ij} p_m$\\
& $\{ q_i , s_j \} = -2i C_{ij} Z.$ \\
\end{tabular}
\end{center}
\be
~
\label{N2D3}
\ee
Note, the notation used in this paper is that of reference \cite{Clark:2002bh}, 
in particular the charge conjugation matrix and the $2+1$ (D=3) dimensional 
gamma matrices in the appropriate associated representation are given there.

The group element $\Omega$ is now parameterized by the coordinates of the 
centrally extended N=2, D=3 superspace, $x^m (\xi), \phi (\xi), \theta_i (\xi), 
\lambda_i (\xi)$.  The above transformation algebra induces a motion in N=2 
superspace,
\bea
x^{\prime m} &=& x^m + a^m  - i (\bar\zeta \gamma^m \theta + \bar\eta \gamma^m 
\lambda ) -\phi b^m + \epsilon^{mnr} \alpha_n x_r \cr
\phi^\prime &=& \phi + z + (\zeta \gamma^0 \lambda - \theta \gamma^0 \eta ) - b^m 
x_m \cr
\theta^\prime_i &=& \theta_i+\zeta_i - \frac{i}{2} b_m (\gamma^m \lambda )_i - 
i\rho 
\lambda_i  -\frac{i}{2} \alpha_m (\gamma^m \theta )_i \cr
\lambda^\prime_i &=& \lambda_i+\eta_i + \frac{i}{2} b_m (\gamma^m \theta )_i + 
i\rho 
\theta_i  -\frac{i}{2} \alpha_m (\gamma^m \lambda )_i  ,
\label{transN2D3}
\eea
resulting in the linear representation of the N=2, D=3 transformation generators
\begin{center}
\begin{tabular}{ll}
$\delta^p(a) x^m = a^m$ & $\delta^Z(z) x^m = 0$ \\
$\delta^p(a) \phi = 0$ & $\delta^Z(z) \phi = z$ \\
$\delta^p(a) \theta = 0$ &  $\delta^Z(z) \theta = 0$ \\
$\delta^p(a) \lambda = 0$ & $\delta^Z(z) \lambda = 0$ \\
 & \\
$\delta^q(\zeta) x^m = -i\bar\zeta \gamma^m \theta$ & $\delta^s(\eta) x^m = -
i\bar\eta \gamma^m \lambda$ \\
$\delta^q(\zeta) \phi = -\bar\zeta \lambda$ & $\delta^s(\eta) \phi = +\bar\eta 
\theta$ \\
$\delta^q(\zeta) \theta_i = \zeta_i$ & $\delta^s(\eta) \theta_i = 0$ \\
$\delta^q(\zeta) \lambda_i = 0$ & $\delta^s(\eta) \lambda_i = \eta_i$ \\
 & \\
$\delta^M(\alpha) x^m = \epsilon^{mnr} \alpha_n x_r$ & $\delta^K(b) x^m = -b^m 
\phi$  \\
$\delta^M(\alpha) \phi = 0$ & $\delta^K(b) \phi = -b^m x_m$ \\
$\delta^M(\alpha) \theta_i = -\frac{i}{2}\alpha_m (\gamma^m \theta)_i$ & 
$\delta^K(b) \theta_i = +\frac{i}{2}b_m (\gamma^m \lambda)_i$ \\
$\delta^M(\alpha) \lambda_i = -\frac{i}{2}\alpha_m (\gamma^m \lambda)_i$ & 
$\delta^K(b) \lambda_i = -\frac{i}{2}b_m (\gamma^m \theta)_i $ \\
 & \\
$\delta^R(\rho) x^m = 0$ & $\delta^R(\rho) \theta_i = -i\rho \lambda_i$ \\
$\delta^R(\rho) \phi = 0$ & $\delta^R(\rho) \lambda_i = +i\rho \theta_i .$ \\
\end{tabular}
\end{center}
\be
~
\label{varN2D3}
\ee
Once again these variations obey the N=2, D=3 SUSY transformation algebra, for 
example
\bea
[\delta^q(\zeta), \delta^q (\zeta^\prime)] &=& 2\delta^p (i\bar\zeta \gamma 
\zeta^\prime ) \cr
[\delta^s(\eta), \delta^s (\eta^\prime)] &=& 2\delta^p (i\bar\eta \gamma 
\eta^\prime ) \cr
[\delta^q(\zeta), \delta^s (\eta)] &=& 2\delta^Z (\bar\zeta \eta) .
\eea
These transformations are precisely of the same form for the total variations of the fields that 
are obtained by means of the coset method.  This is because the same exponential 
representation of the subgroup elements is used as in reference \cite{Clark:2002bh}.

The N=2 SUSY invariant one forms now become
\bea
\Pi^a_m &=& \partial_m x^a +i\bar\theta \gamma^a \partial_m\theta +i\bar\lambda 
\gamma^a \partial_m \lambda  \cr
\Pi^3_m &=& \partial_m \phi +\bar\theta \partial_m\lambda -\bar\lambda 
\partial_m \theta .
\label{1formsN2D3}
\eea
The induced metric on the world volume, requiring D=4 Lorentz invariance 
(specifically under the broken $K^m$ transformations), now is given by
\be
g_{mn} = \Pi^a_m \eta_{ab}\Pi^b_n - \Pi^3_m \Pi^3_n .
\label{metric1}
\ee
Equivalently, equations (\ref{1formsN2D3}) and (\ref{metric1}) are obtained by substituting the expressions
for $\theta^\alpha$ and $\bar\theta_{\dot\alpha}$ in terms of $\theta_i$ and $\lambda_i$ (i.e. $\theta^\alpha Q_\alpha + \bar\theta_{\dot\alpha} \bar{Q}^{\dot\alpha} = \bar\theta_i q_i + \bar\lambda_i s_i $),
\be
\pmatrix{
\theta^\alpha \cr
\bar\theta_{\dot\alpha}\cr} = \frac{1}{2} e^{i\frac{\pi}{4}} \pmatrix{
\sigma_z\sigma^{\rm T} & -i\sigma_z\sigma^{\rm T} \cr
-i\sigma^{\rm T} & -\sigma^{\rm T}\cr} \pmatrix{\theta_i \cr
\lambda_i \cr} ,
\ee
into equations (\ref{1formsN1D4}) and (\ref{metricD4}).  The N=1, D=4 super-Poincar\'{e} invariant action is as previously
\be
\Gamma = -\sigma \int d^3 \xi \sqrt{\det{g_{mn}}} .
\ee

Introducing the Akulov-Volkov induced dreibein in this general gauge
\be
\hat{e}_m^{~a} = \Pi_m^a = \partial_m x^{a} +i\bar\theta \gamma^a \partial_m 
\theta +i\bar\lambda \gamma^a \partial_m \lambda ,
\label{AVdreibeingeneralgauge}
\ee
the induced metric then takes the form
\be
g_{mn} = \hat{e}_m^{~a} n_{ab} \hat{e}_n^{~b} ,
\label{metric2}
\ee
with the Nambu-Goto metric $n_{ab}$ given by
\bea
n_{ab} &=& \eta_{ab} -\hat{e}_a^{-1m} \Pi^3_m \hat{e}_b^{-1n} \Pi_n^3 \cr
 &=& \eta_{ab} -\left\lbrack \hat{\cal D}_a \phi -\hat{\cal D}_a \bar\theta 
\lambda + 
\bar\theta \hat{\cal D}_a \lambda \right\rbrack
\left\lbrack \hat{\cal D}_b \phi -\hat{\cal D}_b \bar\theta \lambda + \bar\theta 
\hat{\cal D}_b \lambda \right\rbrack  ,
\label{NGmetric}
\eea
where the Akulov-Volkov partially covariant derivative, $\hat{\cal D}_a$, and the partially 
covariant derivative of $\phi$, $\hat\nabla_a\phi$, are, respectively,
\bea
\hat{\cal D}_a &\equiv& \hat{e}_a^{-1m} \partial_m \cr
\hat\nabla_a\phi &\equiv& \hat{e}_a^{-1 p} \Pi_p^{~3} = \hat{\cal D}_a \phi 
+\bar\theta \hat{\cal D}_a \lambda - \hat{\cal D}_a \bar\theta \lambda = 
\hat{e}_a^{-1m}(\partial_m \phi +\bar\theta 
{\stackrel{\leftrightarrow}{\partial}}_m \lambda ).
\eea
The determinant of $n_{ab}$ is simply 
\be
\det{n} = 1-(\hat\nabla_a \phi)^2 = 1-\left\lbrack \hat{\cal D}_a \phi -\hat{\cal D}_a \bar\theta \lambda + 
\bar\theta \hat{\cal D}_a \lambda \right\rbrack^2 .
\ee
Hence, taking the square root of this determinant, the action for the non-BPS 
p=2 brane embedded in N=1, D=4 superspace is obtained 
\be
\Gamma = -\sigma \int d^3 \xi \det{\hat{e}} \sqrt{1- \left({\hat{\cal D}}_a \phi 
+ \bar\theta {\hat{\cal D}}_a \lambda -  {\hat{\cal D}}_a \bar\theta \lambda 
\right)^2} .
\label{NGAV1}
\ee
Exploiting the world volume reparametrization invariance in order to choose the 
static gauge, $\xi^m = x^m$, all the world volume fields become 
functions of $x^m$ and the Akulov-Volkov dreibein becomes that of \cite{Clark:2002bh}
\be
\hat{e}_m^{~a} = \Pi_m^a = \delta_m^{~a} +i\bar\theta \gamma^a \partial_m \theta 
+i\bar\lambda \gamma^a \partial_m \lambda .
\ee
The action reproduces that found in \cite{Clark:2002bh}, equation 
(\ref{CNtVAVNG}) above, by means of the coset method.

Once in the static gauge the action is no longer invariant under the original 
symmetry transformations since now $x^m = \xi^m$.  A compensating general 
coordinate transformation must be made on the fields so that when applied to the 
$x^m$ coordinates they are invariant and so remain in the static gauge.  Hence 
intrinsic variations of the fields, $\bar\delta \varphi$, are defined with a 
compensating coordinate transforming factor, denoted as $C^m$,
\bea
\bar\delta x^m &=& \delta x^m - C^l \partial_l x^m \cr 
\bar\delta \phi &=& \delta \phi - C^l \partial_l \phi \cr 
\bar\delta \theta &=& \delta \theta - C^l \partial_l \theta \cr 
\bar\delta \lambda &=& \delta \lambda - C^l \partial_l \lambda .
\eea
In particular $x^m$ must remain in the static gauge, hence it is required that 
$\bar\delta x^m =0$, thus the compensating factor is determined for each super-Poincar\'{e} 
transformation $C^m = \delta x^m$.  Consequently it is obtained that
\bea
C^m(a) &=& a^m \cr
C^m(z) &=& 0 \cr
C^m(\zeta) &=& -i \bar\zeta \gamma^m \theta \cr
C^m(\eta) &=& -i \bar\eta \gamma^m \lambda \cr
C^m(\alpha) &=& \epsilon^{mnr} \alpha_n x_r \cr
C^m(b) &=& -b^m \phi \cr
C^m(\rho) &=& 0 .
\eea
These compensating terms reproduce the intrinsic variations of the fields 
obtained in reference \cite{Clark:2002bh} by means of the coset method.  The 
compensating factors are the Taylor expansion factors necessary to go from the 
linearly represented total variations of the fields to the nonlinear realizations of the 
intrinsic (Lie) transformations.  The static gauge action is invariant under the 
nonlinearly realized N=1, D=4 super-Poincar\'{e} variations of the $\phi,~\theta,~\lambda$
fields (dropping the bar on the intrinsic variations $\bar\delta \rightarrow 
\delta$)
\begin{center}
\begin{tabular}{ll}
$\delta^p(a) \phi = -a^m \partial_m \phi (x)$ & $\delta^Z(z) \phi = z$ \\
$\delta^p(a) \theta =  -a^m \partial_m \theta  (x)$ & $\delta^Z(z) \theta = 0$ 
\\
$\delta^p(a) \lambda = -a^m \partial_m \lambda (x)$  & $\delta^Z(z) \lambda = 0$ 
\\
 & \\
$\delta^q(\zeta) \phi = -\bar\zeta \lambda +i\bar\zeta \gamma^m \theta 
\partial_m \phi (x)$  & $\delta^s(\eta) \phi = +\bar\eta \theta +i\bar\eta 
\gamma^m \lambda \partial_m \phi (x)$  \\
$\delta^q(\zeta) \theta_i = \zeta_i +i\bar\zeta \gamma^m \theta \partial_m 
\theta_i (x)$ & $\delta^s(\eta) \theta_i = +i\bar\eta \gamma^m \lambda 
\partial_m \theta_i (x)$ \\
$\delta^q(\zeta) \lambda_i = +i\bar\zeta \gamma^m \theta \partial_m \lambda_i 
(x)$ & $\delta^s(\eta) \lambda_i = \eta_i +i\bar\eta \gamma^m \lambda \partial_m 
\lambda_i (x)$  \\
 & \\
$\delta^M(\alpha) \phi = -\epsilon^{mnr} \alpha_n x_r \partial_m \phi (x)$ & 
$\delta^K(b) \phi = -b^m x_m  +\phi b^m \partial_m \phi (x)$  \\
$\delta^M(\alpha) \theta_i = -\frac{i}{2}\alpha_m (\gamma^m \theta)_i -
\epsilon^{mnr} \alpha_n x_r \partial_m \theta_i (x)$  & $\delta^K(b) \theta_i = 
+\frac{i}{2}b_m (\gamma^m \lambda)_i +\phi b^m 
\partial_m \theta_i (x)$  \\
$\delta^M(\alpha) \lambda_i = -\frac{i}{2}\alpha_m (\gamma^m \lambda)_i -
\epsilon^{mnr} \alpha_n x_r \partial_m  \lambda_i (x)$  & $\delta^K(b) \lambda_i 
= -\frac{i}{2}b_m (\gamma^m \theta)_i +\phi b^m 
\partial_m \lambda_i (x)$ \\
 & \\
$\delta^R(\rho) \phi = 0$ & \\
$\delta^R(\rho) \theta_i = -i\rho \lambda_i$ & \\
$\delta^R(\rho) \lambda_i = +i\rho \theta_i$ . & \\
\end{tabular}
\end{center}
\be
~
\ee

In general in D=3 a scalar field and a U(1) gauge field provide equivalent 
descriptions for the single field degree of freedom.  Returning to the action in 
a general gauge, equation (\ref{NGAV1}), this relation can be made explicit by 
first introducing a Lagrange multiplier field $L^a$ in order to define $k_a = 
\hat\nabla_a \phi$ so that the action becomes
\be
\Gamma = - \sigma \int d^3 \xi \det{\hat{e}} \left\{\sqrt{(1-k^2)} + L^a (k_a -
\hat\nabla_a \phi)\right\} ,
\label{NGAV7}
\ee
The action is now a functional of the independent fields $\Gamma = \Gamma 
[\theta, \lambda, \phi, k, L]$.  The $\phi$ equation of motion will imply 
the D=3 Bianchi identity for the U(1) field strength Lagrange multiplier field $L^a$
\be
\frac{\delta \Gamma}{\delta \phi} = -\sigma \partial_m \left(\det{\hat{e}} L^a \hat{e}_a^{-1m} 
\right) =0 .
\ee
Hence the field strength $F^m \equiv \det{\hat{e}}L^a \hat{e}_a^{-1m}$ obeys the D=3 Bianchi 
identity $\partial_m F^m =0$, which has the U(1) gauge potential, $A_m$, 
solution 
\be
F^m = \epsilon^{mnr} \partial_n A_r ,
\ee
that is $F_{mn}= \epsilon_{mnr}F^r =\partial_m A_n -\partial_n A_m$.
At the same time the Bianchi identity allows $\phi$ to be eliminated from the 
action by an integration by parts as seen below.  

The $k^a$ field equation is 
algebraic and allows $k^a$ to be directly eliminated from the action in favor of 
$L^a$ and hence $F^m$
\be
\frac{\delta \Gamma}{\delta k_a} = -\sigma \det{\hat{e}} \left(L^a 
- \frac{k^a}{\sqrt{1-k^2}} \right) =0 .
\ee
Thus it is obtained that 
\be
\frac{k^a}{\sqrt{1-k^2}} = L^a .
\ee
Contracting with $L^a$ yields
\be
L^a k_a = L^2 \sqrt{1-k^2} ,
\ee
while squaring yields the equation
\be
\sqrt{1-k^2} = \frac{1}{\sqrt{1 + L^2}}.
\ee
Substituting these into the p=2 brane action (\ref{NGAV7}), the dual D2-brane Born-
Infeld action is secured
\be
\Gamma = - \sigma \int d^3 \xi  \left\{\det{\hat{e}}\sqrt{1+L^2} - F^m 
\left( \partial_m \phi +\bar\theta \stackrel{\leftrightarrow}{\partial_m} 
\lambda \right) \right\} .
\ee
Substituting the U(1) field strength for $L^a$
\be
L^2 = \frac{1}{(\det{\hat{e}})^2}(F^m \hat{e}_m^{~a})\eta_{ab} (F^n \hat{e}_n^{~b})
= F^a \eta_{ab} F^b = F^a F_a  ,
\ee
the action is obtained in the general gauge
\bea
\Gamma &=& -\sigma \int d^3 \xi \left( \det{\hat{e}} \sqrt{1+ F^a F_a}
 -  F^m [\partial_m \phi + \bar\theta 
{\stackrel{\leftrightarrow}{\partial_m}} \lambda ] \right) \cr
 &=& -\sigma \int d^3 \xi \left( \sqrt{\det{(\hat{g}_{mn} + F_{mn})}} -F^m 
(\bar\theta {\stackrel{\leftrightarrow}{\partial}}_m \lambda) \right) ,
\label{BI1}
\eea
where the $\phi$ term has been eliminated by an 
integration by parts and the use of the Bianchi identity $\partial_m F^m =0$, as alluded to above.  
Once again going to the static gauge, the action of \cite{Clark:2002bh}, equation 
(\ref{CNtVBI}) above, is obtained.  Hence the non-BPS p=2 brane 
Nambu-Goto-Akulov-Volkov action (\ref{NGAV1}) obtained by the Green-Schwarz method and the coset method is dual to the 
space-filling N=2, D=3 non-BPS D2-brane supersymmetric 
Born-Infeld action first obtained in \cite{Clark:2002bh} by the coset method and here by Lagrange multiplier and field equation techniques \cite{Townsend:1995af}.

\pagebreak

\newsection{The Non-BPS D2-Brane Action And Duality}

Consider a space-filling non-BPS D2-brane in N=2, D=3 superspace directly.  The 
N=2, D=3 superspace coordinates of the brane are $x^m (\xi)$, $\theta_i (\xi)$ 
and $\lambda_i (\xi)$ with $\xi^m$ the world volume coordinates of the brane.  
The isometries of the target space are now those of the non-centrally extended 
N=2, D=3 super Poincar\'e group.  Their algebra is that given in equation 
(\ref{N2D3}) with $Z=0$ and $K^m$ absent.  Likewise the transformations of the 
fields are just those in (\ref{transN2D3}) and (\ref{varN2D3}) without the 
equations for $\phi$ and with $b^m =0$.
Consequently the SUSY invariant building block is the Akulov-Volkov 
dreibein as in equation (\ref{AVdreibeingeneralgauge})
\be
\hat{e}_m^{~a} = \Pi_m^a = \partial_m x^{a} +i\bar\theta \gamma^a \partial_m 
\theta +i\bar\lambda \gamma^a \partial_m \lambda 
\ee
with the associated invariant Akulov-Volkov metric made with the Minkowski metric $\eta_{ab}$ (not the Nambu-Goto metric $n_{ab}$ as in equation (\ref{metric2}))
\be
\hat{g}_{mn} = \Pi_m^{~a} \eta_{ab} \Pi_n^{~b} = \hat{e}_m^{~a} \eta_{ab} 
\hat{e}_n^{~b} .
\label{D2gmn}
\ee
In addition a non-BPS D2-brane is described by a gauge potential $A_m$ through a 
generalized SUSY invariant field strength ${\cal F}_{mn}= F_{mn} -b_{mn}$ where 
$F_{mn}= \partial_m A_n -\partial_n A_m$ and 
$b_{mn}$ is a world volume 2-form.  The $b_{mn}$ can 
be found by its required property that its SUSY variations are exact, $\delta^{q,s} b = d\beta$.  Following the construction of \cite{Townsend:1995af} and \cite{Sen:1999md}, $b_{mn}$ is given by
\be
b_{mn} = i\left\lbrack \bar\theta \gamma^l \partial_m \theta - \bar\lambda 
\gamma^l 
\partial_m \lambda \right\rbrack  \left\lbrack \partial_n x_l +\frac{i}{2} 
\bar\theta 
\gamma_l \partial_n \theta +\frac{i}{2} \bar\lambda \gamma_l \partial_n \lambda 
\right\rbrack - (m \leftrightarrow n) .
\ee
Applying the SUSY variations to $b_{mn}$ it is found that
\bea
\delta^q (\zeta) b_{mn} &=& \partial_m  \left\lbrack i\bar\zeta \gamma^l \theta 
\partial_n x_l + \frac{1}{2} (\bar\zeta \partial_n \theta ) \bar\theta\theta 
\right\rbrack - \partial_n  \left\lbrack  i\bar\zeta \gamma^l \theta \partial_m 
x_l + \frac{1}{2} (\bar\zeta \partial_m \theta ) \bar\theta\theta  \right\rbrack \cr
\delta^s (\eta) b_{mn} &=& -\partial_m  \left\lbrack  i\bar\eta \gamma^l \lambda 
\partial_n x_l + \frac{1}{2} (\bar\eta \partial_n \lambda )\bar\lambda\lambda 
\right\rbrack + \partial_n  \left\lbrack  i\bar\eta \gamma^l \lambda \partial_m 
x_l + \frac{1}{2} (\bar\eta \partial_m \lambda ) \bar\lambda\lambda  \right\rbrack 
.\cr
 & & 
\eea
Since the variations are exact, ${\cal F}_{mn}$ can be made SUSY invariant, 
$\delta^{ q,s} {\cal F}_{mn}=0$, by canceling the $b_{mn}$ variations with 
those of the gauge field $A_m$
\bea
\delta^q (\zeta) A_m &=& \left\lbrack i\bar\zeta \gamma^l \theta \partial_m x_l 
+ \frac{1}{2} (\bar\zeta \partial_m \theta ) \bar\theta\theta \right\rbrack \cr
\delta^s (\eta) A_m &=& -\left\lbrack i\bar\eta \gamma^l \lambda \partial_m x_l 
+ \frac{1}{2} (\bar\eta \partial_m \lambda ) \bar\lambda\lambda \right\rbrack .
\eea

The N=2, D=3 super Poincar\'e invariant non-BPS D2-brane Born-Infeld action is 
given by, equation (\ref{GSSD2}),
\be
\Gamma = -\sigma \int d^3 \xi \sqrt{\det{(\hat{g}_{mn} + {\cal F}_{mn})}} .
\label{nonBPSD2}
\ee
Recalling equation (\ref{D2gmn}) for the metric, the action becomes, including a 
Lagrange multiplier field, $\Lambda^{mn}$, defining the field strength tensor,
\bea
\Gamma &=& -\sigma \int d^3 \xi  \det{\hat{e}}\left\{\sqrt{\det{(\eta_{ab} + 
\hat{e}_a^{-1m}\hat{e}_b^{-1n}{\cal F}_{mn})}} -\frac{1}{2} \Lambda^{mn} ( 
F_{mn} -(\partial_m A_n -\partial_n A_m ))\right\} \cr
 &=& -\sigma \int d^3 \xi \det{\hat{e}}\left\{ \sqrt{(1+ {\cal F}^a 
{\cal F}_a)} -F^a \Lambda_a + \Lambda^{mn} \partial_m A_n 
\right\}  ,
\label{BI2}
\eea
where for any antisymmetric 2-tensor, $T_{mn}$, the corresponding vector density, $T^m$, is given 
by $T^m = \frac{1}{2} \epsilon^{mnr} T_{nr}$ and vice versa 
$T_{mn} = \epsilon_{mnr} T^r$.  Likewise, $T_m = \frac{1}{2} \epsilon_{mnr} T^{nr}$ and inversely 
$T^{mn} = \epsilon^{mnr} T_r$.  It should be noted that the contravariant and covariant vector densities 
are related according to $T^m = (\det{\hat{e}})^2 \hat{g}^{mn} T_n$.
It is useful to define tangent space tensors, $T_{ab}$ and $T^{ab}$, and their associated tangent space vectors,
$T^a$ and $T_a$,
\bea
T_{ab} &\equiv & \hat{e}_a^{-1m}\hat{e}_b^{-1n} T_{mn} \qquad ; \qquad T^{ab} \equiv 
\hat{e}_m^{a}\hat{e}_n^{b} T^{mn} \cr
T^a &=& \frac{1}{2} \epsilon^{abc} T_{bc} \qquad\qquad ; \qquad 
T_a = \frac{1}{2} \epsilon_{abc} T^{bc} .
\label{tangent}
\eea
The contravariant and covariant tangent space vectors are related as $T^a = \eta^{ab} T_b$.  Ordinary contravariant and covariant world volume vectors are related as usual by $V^m = \hat{g}^{mn} V_n$.
Exploiting the relation
\be
(\det{\hat{e}}) \epsilon_{mnr} = \epsilon_{abc} \hat{e}_m^{~a}\hat{e}_n^{~b}\hat{e}_r^{~c}
\ee
and equation (\ref{tangent}), world volume tensor expressions are readily converted to the corresponding 
tangent space tensor ones, for example, $\frac{1}{2} T^{mn} T_{mn} = T^m T_m = T^a T_a$.

The non-BPS p=2 brane action dual to this non-BPS D2-brane action can be obtained by eliminating the 
now independent field strength tensor.  The field equations for the field strength, $\delta 
\Gamma /\delta F^a =0$, yield an algebraic relation between it and the Lagrange 
multiplier and Goldstino fields
\be
\Lambda_a = \frac{{\cal F}_a}{\sqrt{1+ {\cal F}^b {\cal F}_b}} .
\ee
Squaring this equation yields
\be
\sqrt{1 + {\cal F}^2} = \frac{1}{\sqrt{1-\Lambda^2}} ,
\ee
and contracting the equation with $\Lambda^a$ gives
\be
\Lambda^a F_a = \Lambda^2 \sqrt{1+ {\cal F}^2}  
+ \Lambda^a b_a.
\ee
Substituting these expressions into the action (\ref{BI2}) results in
\be
\Gamma = -\sigma \int d^3 \xi \det{\hat{e}} \left\{ \sqrt{1 -\Lambda^2}
 - \Lambda^a b_a + \frac{1}{2} \Lambda^{mn}(\partial_m A_n -\partial_n A_m 
)\right\} .
\ee
Applying the gauge field equation of motion, $\delta \Gamma / \delta A_n = 0 = 
+\sigma \partial_m ((\det{\hat{e}})\Lambda^{mn})$,
implies that the Lagrange multiplier is given by the curl of a scalar field
$(\det{\hat{e}})\Lambda^{mn} = \epsilon^{mnr} \partial_r \phi$, so that $\Lambda_m = 
\frac{1}{\det{\hat{e}}}\partial_m \phi$, or equivalently
\be
\Lambda_a = \hat{e}_a^{-1m} \partial_m \phi = \hat{{\cal D}}_a \phi .
\ee
Integrating the gauge field terms by parts and setting $\partial_m ((\det{\hat{e}})\Lambda^{mn}) 
=0 $ yields the dual action for a non-BPS p=2 brane embedded in 
N=1, D=4 superspace, equation (\ref{GSSdualp2}),
\be
\Gamma = -\sigma \int d^3 \xi \left[ \det{\hat{e}} \sqrt{ 1 -\hat{{\cal D}}_a 
\phi \hat{{\cal D}}^a \phi } 
- \frac{1}{2} \epsilon^{mnr} b_{mn} \partial_r \phi \right] .
\label{NGAV3}
\ee
\pagebreak

\newsection{Field Redefinitions And Equivalent Actions}

As stated in the introduction, the non-BPS p=2 brane action obtained by the Green-Schwarz and coset
methods, equation (\ref{CNtVAVNG}), differs from the above action, equation (\ref{NGAV3}) ((\ref{GSSdualp2})), obtained by duality from the direct construction of the non-BPS D2-brane action, equation (\ref{nonBPSD2}) ((\ref{GSSD2})).  Or, mutatis mutandis, the Akulov-Volkov-Nambu-Goto action in equation (\ref{NGAV1}) ((\ref{CNtVAVNG})) can itself be obtained as the dual to the non-BPS D2-brane Born-Infeld action in equation (\ref{BI1}) ((\ref{CNtVBI})).  Proceeding analogously to the above, the action (\ref{BI1}) with Lagrange multiplier is introduced
\bea
\Gamma &=& -\sigma \int d^3 \xi \left\{\sqrt{ \det{\left( \hat{g}_{mn} + 
F_{mn}\right)}} - F^m (\bar\theta \stackrel{\leftrightarrow}{\partial_m} \lambda 
)  \right. \cr
 & & \left. \qquad\qquad\qquad\qquad -\frac{1}{2} (\det{\hat{e}}) \Lambda^{mn} ( F_{mn} -(\partial_m A_n -\partial_n A_m )) 
\right\} \cr
 &=& -\sigma \int d^3 \xi \det{\hat{e}}\left\{ \sqrt{(1 + F^a F_a)} - F^a \Lambda_a 
-  F^a (\bar\theta 
\stackrel{\leftrightarrow}{\hat{{\cal D}}_a} \lambda )\right. \cr
 & & \left. \qquad\qquad\qquad\qquad\qquad\qquad + \frac{1}{2} \Lambda^{mn}(\partial_m A_n -\partial_n 
A_m )\right\} .
\label{BI1L}
\eea
Applying the equation of motion for the field strength, 
$\frac{\delta\Gamma}{\delta F_a} = 0$, it is found that
\be
\Lambda_a + (\bar\theta \stackrel{\leftrightarrow}{\hat{{\cal D}}_a} 
\lambda ) = \frac{F_a}{\sqrt{1 + F^b F_b}} .
\ee
Once again, squaring the equation yields
\be
\sqrt{1 + F^2} = \frac{1}{\sqrt{1-(\Lambda + (\bar\theta 
{\stackrel{\leftrightarrow}{\hat{{\cal D}}}} \lambda ))^2}} ,
\ee
while contracting with $(\Lambda_a + (\bar\theta 
{\stackrel{\leftrightarrow}{\hat{{\cal D}}_a}} \lambda ))$ yields
\be
F^a \Lambda_a + F^a (\bar\theta {\stackrel{\leftrightarrow}{\hat{{\cal D}}_a}} \lambda 
)= ( \Lambda + (\bar\theta 
{\stackrel{\leftrightarrow}{\hat{{\cal D}}}} \lambda ))^2 \sqrt{1 + F^2} .
\ee
Substituting these into equation (\ref{BI1L}), the action becomes
\be
\Gamma = -\sigma \int d^3 \xi  \det{\hat{e}}\left\{\sqrt{1 -(\Lambda + 
(\bar\theta {\stackrel{\leftrightarrow}{\hat{{\cal D}}}} \lambda 
))^2} + \frac{1}{2} \Lambda^{mn}(\partial_m A_n -\partial_n A_m ) 
\right\} .
\ee
Applying the $A_n$ field equation results in $\Lambda_a = \hat{{\cal D}}_a \phi$ and 
hence the Nambu-Goto-Akulov-Volkov action of equation (\ref{NGAV1}) for a 
non-BPS p=2 brane embedded in N=1, D=4 superspace is obtained
\be
\Gamma = -\sigma \int d^3 \xi \det{\hat{e}} \sqrt{ 1 - \hat{\nabla}^a \phi 
\hat{\nabla}_a \phi }.
\ee

Since these actions describe the dynamics of the same extended objects, either
a non-BPS p=2 brane or a non-BPS D2-brane, there is an equivalence relation between them.
Returning to the direct Green-Schwarz form of the D2-brane action, equation (\ref{nonBPSD2}), 
a field redefinition for the field strength tensor in equation (\ref{BI2}), now 
denoted with a prime, $F_{mn}^\prime = F_{mn} + y_{mn}$, is required in order to 
express the action in the alternate form of equation (\ref{BI1L}).  The field redefinition 
$y_{mn}$ is determined by requiring the Lagrangian of equation (\ref{BI2}) to be 
equal to that in equation (\ref{BI1L}) 
\bea
\sqrt{\det{(\hat{g}_{mn} + {\cal F}_{mn}^\prime)}} &-& \frac{1}{2}(\det{\hat{e}}) \Lambda^{mn} 
F_{mn}^\prime \cr
 &=& \sqrt{ \det{\left( \hat{g}_{mn} + 
F_{mn}\right)}} - F^m (\bar\theta \stackrel{\leftrightarrow}{\partial_m} 
\lambda )  -\frac{1}{2}(\det{\hat{e}}) \Lambda^{mn} F_{mn} . 
\eea
This yields the equation for $y_{mn}$
\be
\sqrt{(1+ F^2)} \left\lbrack \sqrt{ 1 + \frac{(y-b)^2 + 2 F(y-b)}{1 
+ F^2}} -1 \right\rbrack
= y^a \Lambda_a  - F^a (\bar\theta {\stackrel{\leftrightarrow}{\hat{\cal 
D}}}_a \lambda ).
\ee
Isolating the square root containing $y^a$ and squaring the equation leads to the 
vanishing of
a quadratic form in $y^a$
\be
y^a A_{ab} y^b + 2 B_a y^a + C = 0 ,
\ee
with the coefficients given by
\bea
A_{ab} &=& \eta_{ab} - \Lambda_a \Lambda_b \cr
B_a &=& (F_a - b_a ) + ((F \cdot J) - {\cal L}) \Lambda_a  \cr
C &=& b^2 -2 (F \cdot b) + 2 {\cal L} (F\cdot J) - (F\cdot J)^2  ,
\eea
where $J_a = (\bar\theta {\stackrel{\leftrightarrow}{\hat{\cal D}}}_a \lambda )$ and 
${\cal L} = \sqrt{ 1+ F^2} .$

The solution to this equation is a surface of the general form
\be
y^a = \sqrt{(BA^{-1} B - C)}~ A^{-\frac{1}{2}a}_{~~~~~b} ~u^b - A^{-1a}_{~~~~b} B^b 
,
\ee
where $u^a$ is an arbitrary time-like unit vector, $ u^a u_a =1$, and 
\bea
A^{-1ab} &=& \eta^{ab} + 
\frac{1}{\lbrack 1 -\Lambda^2 \rbrack} \Lambda^a \Lambda^b  \cr
A^{-\frac{1}{2}ab} &=& \eta^{ab} + 
\frac{1}{\Lambda^2} \left\lbrack 1 - 
\frac{1}{\sqrt{1 -\Lambda^2}}\right\rbrack \Lambda^a \Lambda^b 
\eea
so that
\bea
(BA^{-1}B - C) &=& \frac{1}{\lbrack 1 -\Lambda^2 \rbrack}(\Lambda\cdot B)^2 + 
\Lambda^2 \lbrack (F\cdot J) -{\cal L}\rbrack^2 \cr
 & & \cr
 & &\qquad + F^2 - 2{\cal L} (F\cdot J) + (F\cdot J)^2 \cr 
 & & \qquad\qquad + 2 (\Lambda\cdot F -\Lambda\cdot b)\lbrack (F\cdot J) -
{\cal L}\rbrack .
\eea
In the purely bosonic case, when the Goldstinos are absent, the action reduces 
to the pure Born-Infeld action in both cases.  Hence the unit vector $u^a$ is 
fixed by requiring no field redefinition in the case of $\theta_i =0=\lambda_i$.
Thus $y_{mn}\vert_{\theta =0 =\lambda} = 0$ is an equation for $u^a$ 
\be
u = \frac{1}{\sqrt{BA^{-1}B}} A^{-\frac{1}{2}}B \vert_{\theta = 0 = \lambda} ,
\ee
where $C\vert_{\theta =0 =\lambda} =0$.  With zero subscripts denoting the value 
of the quantities at $\theta_i 
=0=\lambda_i$, the solution for the required field redefinition is found
\be
y = \sqrt{1 - \frac{C}{BA^{-1}B}} \frac{\sqrt{BA^{-1}B}}{\sqrt{B_0 A_0^{-1} 
B_0}}A^{-\frac{1}{2}}
A_0^{-\frac{1}{2}} B_0 -A^{-1} B ,
\label{sol}
\ee
with
\bea
B_0 A_0^{-1} B_0 &=& \left\lbrack {\frac{1}{\lbrack 1  -\Lambda^2 \rbrack}}
\lbrack (F\cdot\Lambda)-{\cal L}\Lambda^2 \rbrack 
\right. \cr
 & & \qquad \left. + \Lambda^2 {\cal L}^2  + F^2 - 2{\cal L} (\Lambda\cdot F) 
\right\rbrack\vert_{\theta = 0 = \lambda}.
\eea

The non-BPS p=2 brane action, equation (\ref{NGAV3}), obtained by duality from the non-BPS D2-brane action, 
equation (\ref{nonBPSD2}), also can be 
shown to be equivalent to the Green-Schwarz method and coset 
method action, equation (\ref{NGAV1}), by a field redefinition directly.  
Consider the p=2 brane action of equation (\ref{NGAV3}), including a Lagrange 
multiplier vector density $L^m$ to define the vector field $v_m^\prime = \partial_m \phi$,
\be
\Gamma = -\sigma \int d^3 \xi  \left\lbrack  \det{\hat{e}} \sqrt{1- \hat{e}_a^{-
1m}v_m^\prime \eta^{ab} 
\hat{e}_b^{-1n}v_n^\prime} - \frac{1}{2} \epsilon^{mnr} b_{mn} v_r^\prime -L^m 
\lbrack v_m^\prime -\partial_m \phi \rbrack \right\rbrack .
\label{NGAV4}
\ee
The vector field can be redefined $v_m^\prime = v_m +y_m$ so that this action 
becomes that of equation (\ref{NGAV1}), including a Lagrange multiplier,
\be
\Gamma = -\sigma \int d^3 \xi  \left\lbrack  \det{\hat{e}} \sqrt{1- \hat{e}_a^{-
1m}v_m \eta^{ab} 
\hat{e}_b^{-1n}v_n}  -L^m \lbrack v_m -(\partial_m \phi +(\bar\theta 
{\stackrel{\leftrightarrow}{\partial}}_m \lambda)) \rbrack \right\rbrack .
\label{NGAV5}
\ee
Setting the Lagrangians equal implies the equation for the required field 
redefinition $y_m$
\be
\det{\hat{e}}\sqrt{1-(v+y)^2} -\frac{1}{2} \epsilon^{mnr} b_{mn} 
(v_r +y_r) -L^m (y_m +J_m) = \det{\hat{e}}\sqrt{1- v^2} ,
\ee
where $J_m \equiv (\bar\theta {\stackrel{\leftrightarrow}{\partial}}_m \lambda 
)$.  As earlier, isolating the square root involving the vector $y_m$ and squaring 
yields an equation for the quadratic form in $y$
\be
y_m A^{mn} y_n +2 B^m y_m +C = 0 ,
\ee
with
\bea
A^{mn} &=& (\det{\hat{e}})^2 \hat{g}^{mn} + (b+L)^m (b+L)^n \cr
B^m &=& (\det{\hat{e}})^2 \hat{g}^{mn} v_n +\left\lbrack (b^n v_n) + (L^n J_n) +{\cal L} 
\right\rbrack (b+L)^m  \cr
C &=& \left\lbrack 2{\cal L} + (b^m v_m) + (L^m J_m)\right\rbrack \lbrack (b^n v_n) 
+ (L^n J_n)\rbrack  ,
\label{coeff}
\eea
where now ${\cal L} = \det{\hat{e}}\sqrt{1 - v^2}$, $b^m = \frac{1}{2} \epsilon^{mnr} b_{nr}$
and recall $\hat{g}_{mn} = \hat{e}_m^{~a} \eta_{ab} \hat{e}_n^{~b}$ so that $\hat{g}^{mn} = \hat{e}_a^{-1m} \eta^{ab} \hat{e}_b^{-1n}$ with $\hat{g}^{mn} \hat{g}_{nr} = \delta^m_{~r}.$
Analogous to the previous case, the solution is given by the surface
\be
y_m = \sqrt{BA^{-1} B - C} A^{-\frac{1}{2}}_{mn} u^n - A^{-1}_{mn} B^n .
\ee
The actions are already identical in the bosonic case, hence the field redefinition must vanish for 
$\theta = 0 = \lambda$.  This defines the time-like unit vector $u^m$ to be
\be
u = \frac{A^{-\frac{1}{2}} B}{\sqrt{BA^{-1}B}}\vert_{\theta = 0 = \lambda}
 = \frac{A_0^{-\frac{1}{2}} B_0}{\sqrt{B_0A_0^{-1}B_0}} .
\ee
The final form of the field redefinition is as in equation (\ref{sol}) along with equation (\ref{coeff}) and
\bea
A_0^{mn} &=& A^{mn}\vert_{\theta = 0 = \lambda} = (\det{\hat{e}_0})^2 
\hat{g}_0^{mn} +L^m L^n \cr
B_0^m  &=& B^m \vert_{\theta = 0 = \lambda} = (\det{\hat{e}_0})^2 \hat{g}_0^{mn} v_n + {\cal L}_0 L^m \cr
C_0 &=& C \vert_{\theta = 0 = \lambda} = 0 .
\eea
Also additional field redefinitions are possible which include changes in the 
Lagrange multiplier field as well.  Once again these lead to a quadratic form in 
$y_m$ equal to zero with modified coefficients.

The Akulov-Volkov-Nambu-Goto action in equation (\ref{NGAV1}) ((\ref{CNtVAVNG})) for a non-BPS p=2 brane embedded in N=1, 
D=4 superspace was the same whether obtained by means of the non-BPS Green-Schwarz 
method or the coset method.  This action was shown to be dual to the 
supersymmetric Born-Infeld action, equation (\ref{BI1}) ((\ref{CNtVBI})), for a space-filling non-BPS D2-brane in N=2, 
D=3 superspace, which was first obtained by the coset method in \cite{Clark:2002bh}.  
Applying the Green-Schwarz methods of Sen \cite{Sen:1999md} to the case at hand, 
another form of the action for the non-BPS D2-brane was obtained, equation 
(\ref{nonBPSD2}) ((\ref{GSSD2})).  The dual non-BPS p=2 brane action to this, equation (\ref{NGAV3}) ((\ref{GSSdualp2})), had a form different from the non-BPS Green-Schwarz method and coset method action of equation 
(\ref{NGAV1}) ((\ref{CNtVAVNG})), as it should.  In each case the required field redefinition 
was found to show that each non-BPS p=2 brane action was equivalent and that each non-BPS
D2-brane action was equivalent.  Finally, coset construction techniques can be readily applied to the 
case of a non-BPS brane in arbitrary space-time dimensions, in particular for ten or eleven dimensions relevant for string theory.
\bigskip
~\\
\noindent
MN would like to thank Koji Hashimoto, Kenji Nagami, Norisuke Sakai and Cosmas Zachos for interesting discussions and useful comments.  The work of TEC and MN was supported in part by the U.S. Department of Energy under grant DE-FG02-91ER40681 (Task B). 

\newpage

\newpage
\begin{thebibliography}{99}
\bibitem{GS}
M.B. Green and J.H. Schwarz, Phys. Lett. {\bf 136B}, 367 (1984);
Nucl. Phys. {\bf B243} 285 (1984);
J.~Hughes, J.~Liu and J.~Polchinski,
Phys.\ Lett.\ B {\bf 180}, 370 (1986);
E. Bergshoeff, E. Sezgin and P.K. Townsend, Phys. Lett. {\bf B189} 75 (1987);
Ann. Phys. {\bf 185} 330 (1988); A. Achucarro, J.M. Evans, P.K. Townsend and 
D.L. Wiltshire, Phys. Lett. {\bf B198} 441 (1987);
M.J. Duff, J. Class. Quant. Grav. {\bf 5} 189 (1988).
\bibitem{Siegel}
W. Siegel, Phys. Lett. {\bf 128B} 397 (1983).
\bibitem{Townsend:1995af}
P.~K.~Townsend,
Phys.\ Lett.\ B {\bf 373}, 68 (1996)
[arXiv:hep-th/9512062];
A.~A.~Tseytlin,
Nucl.\ Phys.\ B {\bf 469}, 51 (1996)
[arXiv:hep-th/9602064];
M.~Aganagic, C.~Popescu and J.~H.~Schwarz,
Phys.\ Lett.\ B {\bf 393}, 311 (1997)
[arXiv:hep-th/9610249];
M.~Aganagic, C.~Popescu and J.~H.~Schwarz,
Nucl.\ Phys.\ B {\bf 495}, 99 (1997)
[arXiv:hep-th/9612080];
M.~Aganagic, J.~Park, C.~Popescu and J.~H.~Schwarz,
Nucl.\ Phys.\ B {\bf 496}, 215 (1997)
[arXiv:hep-th/9702133];
C.~Schmidhuber,
Nucl.\ Phys.\ B {\bf 467}, 146 (1996)
[arXiv:hep-th/9601003];
S.~P.~de Alwis and K.~Sato,
Phys.\ Rev.\ D {\bf 53}, 7187 (1996)
[arXiv:hep-th/9601167];
M.~Cederwall, A.~von Gussich, B.~E.~Nilsson, P.~Sundell and A.~Westerberg,
Nucl.\ Phys.\ B {\bf 490}, 179 (1997)
[arXiv:hep-th/9611159];
E.~Bergshoeff and P.~K.~Townsend,
Nucl.\ Phys.\ B {\bf 490}, 145 (1997)
[arXiv:hep-th/9611173];
Y.~Lozano,
Phys.\ Lett.\ B {\bf 399}, 233 (1997)
[arXiv:hep-th/9701186].
\bibitem{Duff:1992hu}
M.~J.~Duff and J.~X.~Lu,
Nucl.\ Phys.\ B {\bf 390}, 276 (1993)
[arXiv:hep-th/9207060].
\bibitem{Chibisov:1997rc}
B.~Chibisov and M.~A.~Shifman,
Phys.\ Rev.\ D {\bf 56}, 7990 (1997)
[Erratum-ibid.\ D {\bf 58}, 109901 (1998)]
[arXiv:hep-th/9706141].
\bibitem{Sen:1999md}
A.~Sen,
JHEP {\bf 9910}, 008 (1999)
[arXiv:hep-th/9909062].
\bibitem{More}
A.~Sen,
JHEP {\bf 9808}, 010 (1998)
[arXiv:hep-th/9805019];
O.~Bergman and M.~R.~Gaberdiel,
Phys.\ Lett.\ B {\bf 441}, 133 (1998)
[arXiv:hep-th/9806155].
\bibitem{Tachyon}
M.~R.~Garousi,
Nucl.\ Phys.\ B {\bf 584}, 284 (2000)
[arXiv:hep-th/0003122].
E.~A.~Bergshoeff, M.~de Roo, T.~C.~de Wit, E.~Eyras and S.~Panda,
JHEP {\bf 0005}, 009 (2000)
[arXiv:hep-th/0003221].
A.~Sen,
Phys.\ Rev.\ D {\bf 68}, 066008 (2003)
[arXiv:hep-th/0303057].
\bibitem{Volkov:jx}
D.~V.~Volkov and V.~P.~Akulov,
JETP Lett.\  {\bf 16}, 438 (1972).
\bibitem{Coleman:sm}
S.~R.~Coleman, J.~Wess and B.~Zumino,
Phys.\ Rev.\  {\bf 177}, 2239 (1969);
C.~G.~Callan, S.~R.~Coleman, J.~Wess and B.~Zumino,
Phys.\ Rev.\  {\bf 177}, 2247 (1969);
D.~V.~Volkov, Sov.\ J.\ Particles and Nuclei {\bf 4}, 3
(1973); V.~I.~Ogievetsky, Proceedings of the X-th Winter School of
Theoretical Physics in Karpacz, vol. 1, p. 227 (Wroclaw, 1974).
\bibitem{Ivanov:1999fw}
E.~Ivanov and S.~Krivonos,
Phys.\ Lett.\ B {\bf 453} (1999) 237
[arXiv:hep-th/9901003];
S.~Bellucci, E.~Ivanov and S.~Krivonos,
Phys.\ Lett.\ B {\bf 482}, 233 (2000)
[arXiv:hep-th/0003273];
S.~Bellucci, E.~Ivanov and S.~Krivonos,
Nucl.\ Phys.\ Proc.\ Suppl.\  {\bf 102}, 26 (2001)
[arXiv:hep-th/0103136].
J.~Hughes, J.~Liu and J.~Polchinski,
Phys.\ Lett.\ B {\bf 180}, 370 (1986).
J.~Hughes and J.~Polchinski,
Nucl.\ Phys.\ B {\bf 278}, 147 (1986).
J.~Bagger and A.~Galperin,
Phys.\ Lett.\ B {\bf 336}, 25 (1994)
[arXiv:hep-th/9406217];
J.~Bagger and A.~Galperin,
Phys.\ Lett.\ B {\bf 412}, 296 (1997)
[arXiv:hep-th/9707061].
\bibitem{Bagger:1996wp}
J.~Bagger and A.~Galperin,
Phys.\ Rev.\ D {\bf 55}, 1091 (1997)
[arXiv:hep-th/9608177];
J.~Bagger and A.~Galperin,
arXiv:hep-th/9810109;
M.~Rocek and A.~A.~Tseytlin,
Phys.\ Rev.\ D {\bf 59}, 106001 (1999)
[arXiv:hep-th/9811232].
E.~Ivanov,
Theor.\ Math.\ Phys.\  {\bf 129}, 1543 (2001)
[Teor.\ Mat.\ Fiz.\  {\bf 129}, 278 (2001)]
[arXiv:hep-th/0105210];
E.~Ivanov,
AIP Conference Proceedings, {\bf 589}, 61 (2001);
E.~Ivanov,
arXiv:hep-th/0002204.
\bibitem{Hara:2000sg}
T.~Hara and T.~Yoneya,
Nucl.\ Phys.\ B {\bf 602}, 499 (2001)
[arXiv:hep-th/0010173].
\bibitem{Clark:2002bh}
T.~E.~Clark, M.~Nitta and T.~ter Veldhuis,
Phys.\ Rev.\ D {\bf 67}, 085026 (2003)
[arXiv:hep-th/0208184];
arXiv:hep-th/0209142, to appear in Phys.\ Rev.\ D.
\bibitem{Ivanov:1975zq}
E.~A.~Ivanov and V.~I.~Ogievetsky,
Teor.\ Mat.\ Fiz.\  {\bf 25}, 164 (1975).
\bibitem{Nambu:1974zg}
Y.~Nambu,
Phys.\ Rev.\ D {\bf 10}, 4262 (1974);
T.~Goto,
Prog.\ Theor.\ Phys.\  {\bf 46}, 1560 (1971).
\bibitem{Born:gh}
M.~Born and L.~Infeld,
Proc.\ Roy.\ Soc.\ Lond.\ A {\bf 144}, 425 (1934);
P.~A.~Dirac,
Proc.\ Roy.\ Soc.\ Lond.\ A {\bf 268}, 57 (1962).
\end{thebibliography}
\end{document}